\documentclass[conference]{IEEEtran}
\usepackage[utf8]{inputenc}

\usepackage{graphicx}
\usepackage{caption,tabularx,booktabs}
\usepackage{subcaption}
\usepackage{mathptmx}
\usepackage[intlimits,namelimits]{amsmath}
\usepackage[english]{babel}
\usepackage{url}
\usepackage{array}
\usepackage{todonotes}
\usepackage{glossaries}
\usepackage{soul}
\usepackage{amsmath}
\usepackage{csquotes}

\newacronym{3GPP}{3GPP}{3rd Generation Partnership Project}
\newacronym{5G}{5G}{Fifth Generation}
\newacronym{API}{API}{application programming interface}
\newacronym{AR}{AR}{augmented reality}
\newacronym{CapEx}{CapEx}{capital expenditure}
\newacronym{DevOps}{DevOps}{Development Operations}
\newacronym{DSP}{DSP}{digital signal processor}
\newacronym{E2E}{E2E}{end-to-end}
\newacronym{LTE}{LTE}{Long Term Evolution}
\newacronym{EC}{EC}{edge computing}
\newacronym{MEC}{MEC}{Multi-Access Edge Computing}
\newacronym{MNO}{MNO}{mobile network operator}
\newacronym{NFV}{NFV}{network function virtualization}
\newacronym{NIC}{NIC}{network interface controller}
\newacronym{OpEx}{OpEx}{operating expenditure}
\newacronym{OSS}{OSS}{open-source software}
\newacronym{OTA}{OTA}{over-the-air}
\newacronym{QoS}{QoS}{quality of service}
\newacronym{RAN}{RAN}{radio access network}
\newacronym{SDN}{SDN}{software defined network}
\newacronym{SIM}{SIM}{subscriber identity module}
\newacronym{UE}{UE}{user equipment}
\newacronym{VOR}{VOR}{Vestibulo-Ocular Reflex}
\newacronym{VR}{VR}{virtual reality}
\newacronym{SDR}{SDR}{software-defined radio}
\newacronym{PaaS}{PaaS}{Platform-as-a-Service}
\newacronym{HMD}{HMD}{head-mounted display}
\newacronym{IoT}{IoT}{Internet of Things}
\newacronym{GPU}{GPU}{graphics processing unit}
\newacronym{IFTTT}{IFTTT}{if this, then that}
\newacronym{LIDAR}{LIDAR}{light detection and ranging}
\newacronym{SLAM}{SLAM}{simultaneous localization and mapping}
\newacronym{COTS}{COTS}{commercial off-the-shelf}
\newacronym{HCI}{HCI}{human-computer interaction}
\newacronym{AI}{AI}{artificial intelligence}
\newacronym{LBS}{LBS}{location-based service}
\newacronym{DoF}{DoF}{depth of field}
\newacronym{RSS}{RSS}{radio signal strength}
\newacronym{RF}{RF}{radio frequency}
\newacronym{RFID}{RFID}{radio frequency identification}
\newacronym{9DoF}{9DoF}{9 degrees of freedom}
\newacronym{URLLC}{URLLC}{ultra-reliable and low-latency communication}
\newacronym{IMU}{IMU}{intertial measurement unit}
\newacronym{USDZ}{USDZ}{universal scene description zip}
\newacronym{PoC}{PoC}{proof of concept}
\newacronym{SSH}{SSH}{secure shell}
\newacronym{LCD}{LCD}{liquid-crystal display}
\newacronym{GPIO}{GPIO}{general-purpose input/output}
\newacronym{TTS}{TTS}{text-to-speech}
\newacronym{STUN}{STUN}{session traversal utilities for NAT}
\newacronym{P2P}{P2P}{peer to peer}
\newacronym{RSRP}{RSRP}{reference signals received power}
\newacronym{RSRQ}{RSRQ}{received signal reference quality}
\newacronym{UDP}{UDP}{user datagram protocol}
\newacronym{EPC}{EPC}{evolved packet core}
\newacronym{DNS}{DNS}{domain name system}
\newacronym{RTP}{RTP}{real-time transport protocol}
\newacronym{WebRTC}{WebRTC}{web real-time communication}
\newacronym{MME}{MME}{mobility management entity}
\newacronym{SoC}{SoC}{system on chip}
\newacronym{BS}{BS}{base station}

\usepackage{balance}

\newcolumntype{L}[1]{>{\raggedright\let\newline\\\arraybackslash\hspace{0pt}}m{#1}}
\newcolumntype{C}[1]{>{\centering\let\newline\\\arraybackslash\hspace{0pt}}m{#1}}
\newcolumntype{R}[1]{>{\raggedleft\let\newline\\\arraybackslash\hspace{0pt}}m{#1}}

\begin{document}

\title{Orchestrating Service Migration for Low Power MEC-Enabled IoT Devices}

\author{
\IEEEauthorblockN{
Jude Okwuibe\IEEEauthorrefmark{1}, 
Juuso Haavisto\IEEEauthorrefmark{2}, 
Erkki Harjula\IEEEauthorrefmark{1}},
Ijaz Ahmad\IEEEauthorrefmark{1} and
Mika Ylianttila\IEEEauthorrefmark{1}

\IEEEauthorblockA{
\IEEEauthorrefmark{1}Centre for Wireless Communications, University of Oulu, Oulu, Finland\\
\IEEEauthorrefmark{2}Center for Ubiquitous Computing, University of Oulu, Oulu, Finland\\
Email: \{\small \tt first.last\}@oulu.fi}
}

\maketitle

\begin{abstract}
\gls{MEC} is a key enabling technology for \gls{5G} mobile networks. \gls{MEC} facilitates distributed cloud computing capabilities and information technology service environment for applications and services at the edges of mobile networks. This architectural modification serves to reduce congestion, latency, and improve the performance of such edge colocated applications and devices. In this paper, we demonstrate how reactive service migration can be orchestrated for low-power \gls{MEC}-enabled \gls{IoT} devices. Here, we use open-source Kubernetes as container orchestration system. Our demo is based on traditional client-server system from \gls{UE} over \gls{LTE} to the \gls{MEC} server. As the use case scenario, we post-process live video received over \gls{WebRTC}. Next, we integrate orchestration by Kubernetes with S1 handovers, demonstrating \gls{MEC}-based \gls{SDN}. Now, edge applications may reactively follow the \gls{UE} within the \gls{RAN}, expediting low-latency. The collected data is used to analyze the benefits of the low-power \gls{MEC}-enabled \gls{IoT} device scheme, in which \gls{E2E} latency and power requirements of the \gls{UE} are improved. We further discuss the challenges of implementing such schemes and future research directions therein.
\end{abstract}

\section{Introduction}

5G MEC services are envisioned to run as close to the \gls{UE} as possible through a series of optimization techniques. In practice, these applications can be considered as agents which move within the \gls{MNO}'s \gls{RAN}, reserving resources and migrating from one MEC host to another in a dynamic fashion. On the MEC hosts, this sort of functionality can be achieved with an orchestrator. The orchestrator's purpose is to enable resource scheduling and application migration while integrating with various parts of the \gls{RAN}. 

However, this sort of mobile and distributed operation infrastructure poses challenges in software engineering. Considering end-user edge applications running on the \gls{MNO}'s infrastructure, the software has to be designed for mobility. In other words, the developer has to assume the host process can and will migrate between MEC hosts. The two common reasons for such migration are to preserve latency guarantees and to optimize radio resources.

In software engineering, similar challenges have been researched in the context of microservices, which operate in containers managed by an orchestrator. Microservices are documented to address scaling challenges of business components and agile software development with large teams of engineers. Furthermore, by limiting and agreeing on interfaces and on how software components are deployed, e.g., via containers, the operational infrastructure is further simplified. This facilitates maintainability, and system reliability as both the hardware in the datacenter and the software managed by an orchestrator can be mutated as long as it adheres to container runtime specifications.

Now, with 5G, we are introduced to a shift towards \gls{SDN} and \gls{NFV}, which mark an evolution from high \gls{CapEx} \gls{DSP} hardware towards more interoperable, general-purpose computing-based cellular network architecture. Via these changes, the MNOs are offered with ever more affordable means of innovation in offering site-specific services accessible on UEs. That is because the SDN hardware of the teleoperator can double as general-purpose computing hardware. It is envisioned some of this computation capability could be passed on to the \glspl{UE} via \gls{MEC}. In literature, the MEC paradigm is often referred to as the enabler for \gls{AR}, \gls{VR}, autonomous cars, and inter-MNO resource market called network slicing.

This paper is organized as follows. Section \ref{ch:bg} discusses the related paradigms in software and telecommunications research. Section \ref{ch:design} combines aspects introduced in Section \ref{ch:bg} to form an environment on which empirical observations and data collection can be done, facilitating future research. Section \ref{ch:eval} evaluates the reactive system on an \gls{OTA} environment on \gls{LTE}. Finally, the contribution is concluded in Section \ref{ch:concl}.

\section{Background}
\label{ch:bg}

Since our surroundings are getting smarter, it is evident that \gls{IoT} devices will constitute a significant part of the \gls{5G} ecosystem and future mobile networks. Low power and limited computational capabilities characterize most IoT devices. These devices are mostly embedded with wireless sensors, cameras, and actuators of all sorts \cite{bonomi2012fog}. The application areas range from pipeline monitoring to closed-loop control of industrial systems, connected rail, wind farms, smart traffic light systems, orchestration of many autonomous yet coordinated components, to applications in the oil and gas sector. Traditionally, IoT devices in these application areas are remote and need some level of autonomy with regards to power and computational resources. However, recent innovations are seeking to optimize such IoT scenarios by augmenting devices with some computational offloading schemes as well as ambient energy harvesting techniques. For such application areas; latency, mobility management, and location awareness are significant bottlenecks to the effective operation of IoT systems. Thus, the traditional two-tiered approach where IoT devices are set to communicate with a server in some hyperscale datacenter directly will not be an effective solution for such IoT-driven systems \cite{bonomi2012fog}. This emphasizes the need for a multi-tier solution \cite{klas2015fog} with MEC serving as a mid-tier between the IoT device cluster and cloud servers, constituting a so-called MEC-enabled IoT devices. 

\subsection{Low Power MEC-Enabled IoT Devices}

The fundamental value proposition of MEC is to advance low latency, higher bandwidth, and more computational capabilities at the edge of mobile networks closer to the end users. This is the same location where IoT devices are expected to dominate in 5G and future networks. For developers and equipment providers, this opens up a vast potential for innovation towards the IoT applications. For service and content providers, this innovation introduces the need for more flexible and robust orchestration platforms that would coordinate these myriads of devices for resource allocation, data sharing, and service distribution.

Edge computing in general has been widely researched to optimize end-user device resources: cyberforaging, \cite{sharifi2012survey}, grid technology \cite{foster2001anatomy}, computation offloading \cite{fernando2013mobile}, cloudlets \cite{satyanarayanan2009case}, and fog computing \cite{yi2015fog}. Fog computing is mainly aimed at IoT applications that leverage a platform set that collectively assists UEs, while MEC is built on the premise of application-related enhancements with regards to feedback mechanisms, content processing, and information storage\cite{taleb2017multi}. Fog computing extends cloud computing capabilities by moving computation and data storage to the edge of the network, allowing for reduced latency and response delay jitter for applications\cite{bonomi2012fog, yi2015survey}. 

These features are particularly critical for latency-sensitive applications such as gaming and video streaming. More particularly, in an IoT environment where applications and sensor embedded physical devices can be leveraged as fundamental appliances and composed in a mashup style to control development cost and maintenance pressure \cite{wen2017fog}. With the present-day design, there is a common tendency for IoT devices to experience crashes and timing failures from low-sensor battery power, high network latency, and low computational capabilities. 

In all ramifications, orchestration remains the key concept within such distributed systems. It enables the alignment of deployed applications within the business interest of users. However, orchestration, as it is today, is unlikely to serve the needs of future IoT applications, mainly because of the diversity, e.g., configuration, location, reliability, scalability, and security that exists among IoT nodes \cite{wen2017fog}.

\subsection{LTE Handover}

In LTE, handover is generally triggered by a simple event described as;

\begin{equation}
    \mathbf{M}_n> \mathbf{M}_s + \mathbf{HOmargine},
\end{equation}

Where $\mathbf{M}_n$ is the \gls{RSRP} in dBm or \gls{RSRQ} in dB for a neighboring cell. $\mathbf{M}_s$ is \gls{RSRP} or \gls{RSRQ} of the serving cell, whereas $\mathbf{HOmargine}$ is the margin between $\mathbf{M}_n$ and $\mathbf{M}_s$. Each cell can have its own $\mathbf{HOmargine}$ values, and the handover decision, based on equation (1), and is carried out using the parameters in UEs. The \gls{BS} can see if a cell is overloaded, or if a UE is moving from one cell to another, and thus can cause a handover using the values of $\mathbf{M}_n$, $\mathbf{M}_s$, and $\mathbf{HOmargine}$. The X2 interface is used to share information between \glspl{BS}, and enable \glspl{BS} to cause handover for load balancing purposes through resource status response and update messages. The X2 interface is also involved in sharing information related to interference management and has a direct influence on some radio resource management processes in real time. The S1 interface connects \gls{MME} servers in LTE and is used to share load information as well as for load balancing among many \glspl{MME}.

\subsection{Container-based microservice architectures}

In recent years, cloud services have been transforming from monolithic architectures towards microservice architectures, where services are composed of various microservices taking care of some limited set of functions \cite{MicroserviceCloud, MonolithVsMicroservCloud}. This microservice approach brings several benefits over monolithic architectures, including better maintainability, flexibility, scalability, and efficiency, as well as reduced complexity. Since each microservice can be developed, tested, deployed, scaled, operated, and upgraded independently, the microservice model is also very flexible in the geographical distribution of computational tasks. 

As mentioned earlier in this paper, MEC brings new computational tier to cloud computing, between the datacenter and local devices \cite{ETSI_SwForMEC_whitepaper}. By moving some functions from the datacenter to MEC, cloud systems can better serve applications requiring low latency while saving computational and networking resources at core networks and datacenters. MEC and microservice architecture fit together: low latency and data processing services, e.g., filtering and fusions, are beneficial to deploy at MEC. Regarding latency, the roundtrip time between the local and MEC node is low, and for data, less of it needs to be delivered to the public cloud.

Microservice architectures are typically implemented using container technology \cite{MicroservicePerformance2015}. However, unlike the monolithic architectures where the whole system runs inside a single container, here containers enable developing applications in a manner where only one or few processes run inside a single container. Docker containers provide a lightweight, low overhead and fast technology empowering the usage of microservice architectures \cite{DockerIoT}.

Orchestration is a technology for controlling interactions between virtualized components such as containers and taking care of service composition, management, and termination. The most commonly used container orchestration technologies are Docker Swarm, Kubernetes and Mesos, all which provide automated support to, e.g., service discovery, load balancing, and software upgrades \cite{MicroserviceBenchmark}.

In general, the dynamic service deployment following UEs, discussed in the introduction, can be realized using microservice architecture implemented using containers and their dynamical orchestration. In this scenario, the orchestrator would deploy the service instances in optimal locations in MEC hosts near the UE as they move across the RAN. The orchestrator is then used to reschedule resources and migrate the applications while simultaneously integrating with various parts of the \gls{RAN}.

\subsection{Related work}

In \cite{cominardi2018opportunities}, authors discussed the prospects of a joint Edge and Fog orchestration to cope with the vast data volume and low latency requirements of 5G and future networks. In this work, the role of such an orchestration platform was analyzed for different diverse 5G scenarios among which was massive IoT, vehicular communication, multi-access network integration, and localized real-time control. The efficient operation of resource-constrained devices in 5G, which would be mostly IoT devices, is one of the cardinal opportunities presented by the integration of such edge-and-fog orchestration capabilities to 5G. These resource-constrained IoT devices can then rely on the edge resources to execute some of their computationally and power demanding tasks, hence enabling a low-cost design for IoT devices without compromising their needed intelligence and capabilities. 

In \cite{hegyi2016application}, authors discussed application orchestration in mobile edge cloud. Here the focus is on the benefits of mobile edge computing towards IoT deployment and how orchestration and application life cycle management plays out in MEC. Placement of components across several layers of telco network was one of the significant complexities discussed in this report, with regards to the implementation of the model. Other complexities such as computational complexity and inherent stochastic nature of the arising problems were also discussed. Mathematical modeling with constrained multi-objective optimization was presented to provide some simplification for implementing this model in a real-life scenario.

In \cite{chiang2016fog} and \cite{taleb2014lightweight}, authors discussed different mechanisms that can be deployed to ensure scalability for a group of low mobility Machine-Type-Communications (MTC) devices at the edge of the radio access network. These mechanisms are based on group profiling to reduce the amount of signaling and their contents, hence enhancing computation offloading and resource allocation for low-power IoT edge devices. With MEC as one of its key components, such mechanisms can speed-up data delivery with fewer requirements on the network capacity\cite{taleb2017multi}. The work presented in \cite{aazam2014fog} proposed a smart gateway solution for filtering IoT communications through some form of data trimming to reduce unnecessary communication that could burden the core network and the cloud datacenter. Thus making the integration of IoT and cloud computing termed Cloud of Things (CoT) a more practical means of reducing the computational resources and data management needed on different IoT nodes. 

Another work of interest was presented in \cite{sun2016edgeiot}, here an edge IoT architecture called edgeIoT was proposed to handle data streams at the mobile edge to address scalability problems with traditional IoT architectures. Here, instead of transmitting data streams generated from IoT devices to a remote cloud server for analysis, each \gls{BS} is connected to a fog node, and the fog node provides computing resources locally. On top of the fog, nodes would be an SDN-based cellular core designed to facilitate packet forwarding among the fog nodes. Also, a hierarchical fog computing architecture is used on each fog node to provide flexible IoT services without compromising users privacy. This is accomplished by associating each user IoT node with a proxy virtual machine in the cloud to perform data computation and analysis before sending the metadata to the corresponding IoT application virtual machine for a response.    

\section{System Framework}\label{ch:design}

\begin{figure}
  \centering
  \includegraphics[width=0.45\textwidth]{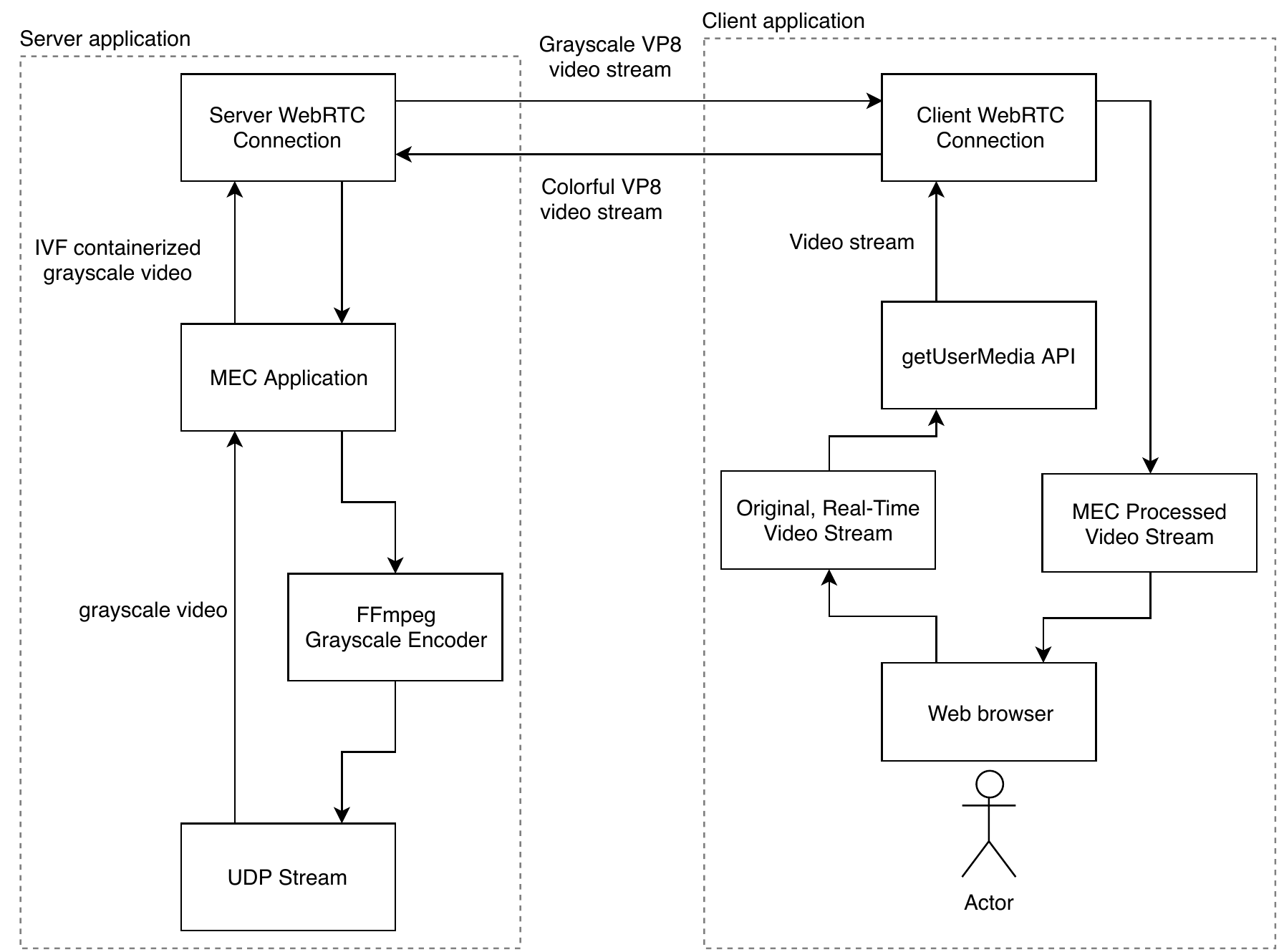}
  \caption{Flowchart of the MEC application logic.}
  \label{fig:mec-app}
\end{figure}

\begin{figure}
  \centering
  \includegraphics[width=0.45\textwidth]{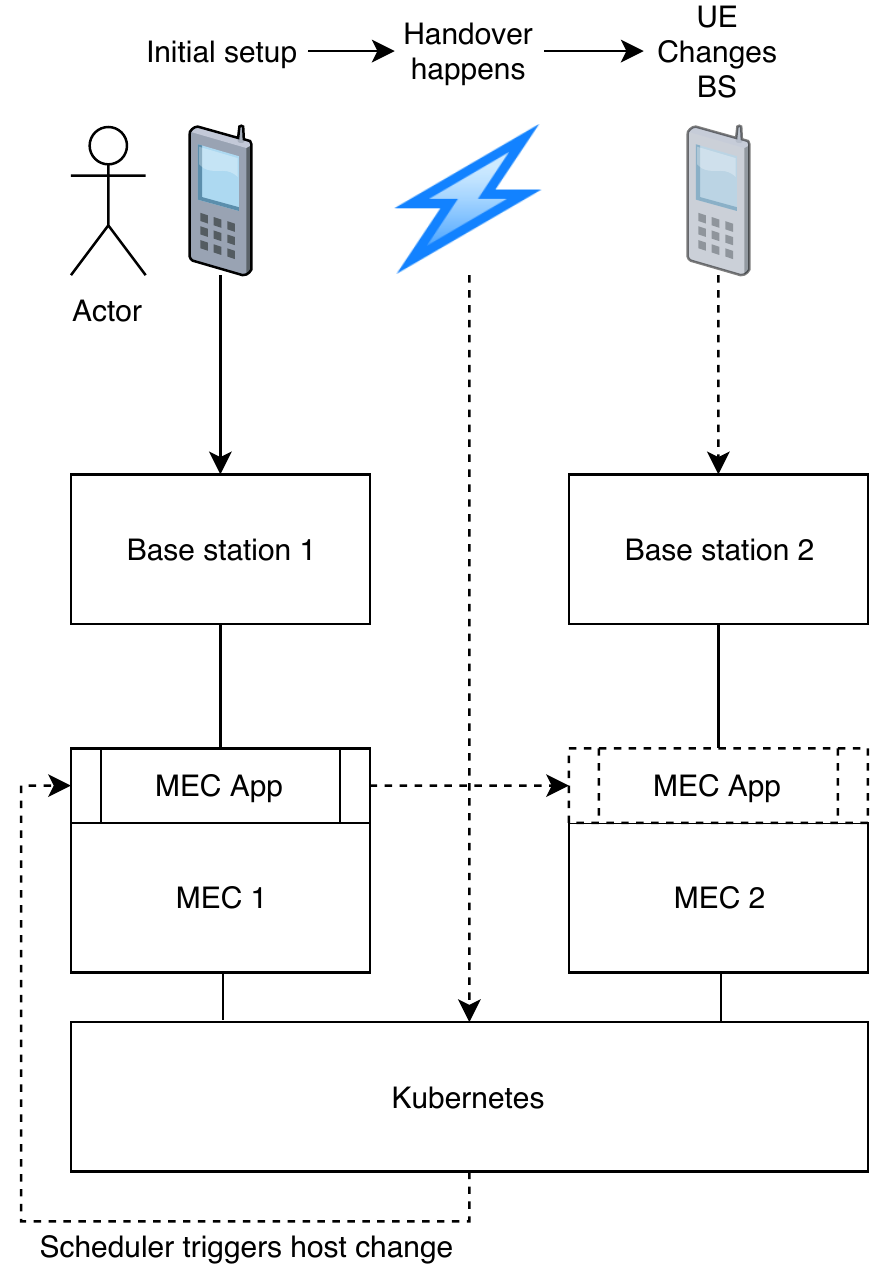}
  \caption{Reactive application migration logic.}
  \label{fig:mec-migration}
\end{figure}

We demonstrate a reactive MEC service migration by presenting a video streaming application. Here, a \gls{MEC} server managed by Kubernetes applies a grayscale filter on an incoming video. A UE captures colorized video stream and runs a web application displaying its original colorized camera stream side-by-side with the \gls{MEC} post-processed one. We then record MEC's end-to-end latency, jitter, and service disruptions in real-time. We present datasets and the source code for the demos and integrations accessible on Github\footnote{https://github.com/toldjuuso/handover2019okwuibe}.

Our environment demonstrated in Fig. \ref{fig:mec-migration}, has two \glspl{BS} and two MEC servers. First, we are assigned to a \gls{BS} and have the MEC application running on MEC 1. Then, we move closer to the second \gls{BS}, which will cause an S1 handover request, perceivable on the \gls{EPC}. Now, the EPC uses the S1 handover request to reactively trigger a Kubernetes service migration to move the edge application from MEC 1 to MEC 2. The migration disrupts the UE application, perceivable by the grayscale video stream stopping and disconnecting. Now, the UE starts polling the MEC cluster to re-establish the video stream. Once Kubernetes has migrated and started the MEC application on the new host, a \gls{DNS} update occurs. This DNS update points the old \gls{MEC} service address, used by the \gls{UE} application, to point to the new host. Now, the connection is re-established, and the grayscale video stream continues. This also marks our demonstration of a reactive MEC service application migration to end.

Our application is an extension to our previous research in Haavisto et al. \cite{haav1906open}, in which an open-source RAN with \gls{DevOps} capabilities is presented. Here, the EPC is part of the same layer three network fabric as the MEC hosts through \verb|flannel|. Because \gls{EPC} handles UE traffic, it can route requests between its intranetwork Kubernetes services and the UEs. As a new contribution, we introduce (1) an application on the environment, and (2) Kubernetes as an integrated facilitator of \gls{RAN} features, here, for MEC service migration on S1 radio handovers. We use S1 handovers instead of X2. This is because the \gls{EPC}, which also runs the \gls{MME}, can then tap network traffic to integrate MEC application migration via Kubernetes.

\subsection{Design of the Client Application}

To elaborate on the right side of Fig. \ref{fig:mec-app}, the client application uses web browser's \verb|getUserMedia| \gls{API} to access its live camera feed. We then leverage WebRTC to send a \gls{UDP} stream of the camera feed to the MEC. By relying on APIs available on web browsers, we achieve interoperability in our demonstration. That is, as long as the UE can run a modern web browser and has a camera attached to it, the UE can be used for the demo. 

\subsection{Design of the Server Application}

To elaborate on the left side of Fig. \ref{fig:mec-app}, the server application expects a VP8 encoded \gls{RTP} stream over WebRTC from the client application. The server then encapsulates the data to an IVF container, and pipes it to \verb|FFmpeg| \cite{ffmpeg}. \verb|FFmpeg| is then responsible for applying a grayscale filter on the video, and emitting it as VP8 stream over \gls{UDP} on a localhost socket. The server application then reads this local \gls{UDP} socket and creates video samples to send back to the client over WebRTC.

\subsection{Design of Reactive Service Migration}

As mentioned, we use Kubernetes as MEC orchestrator and an open-source NextEPC \cite{nextepc} as the \gls{EPC}. To integrate MEC application migration on handovers, we create a script which scans the standard output of the EPC. Now, when the EPC logs a handover request, the script triggers an automatic service migration to the Kubernetes scheduler. Here, the original MEC application host is first cordoned as unschedulable. Then the Kubernetes pod of the video post-processing service is killed. Kubernetes scheduler then reacts by placing the new pod on the only available server, which is MEC 2. Once the migration is done, CoreDNS \cite{coredns}, our chosen DNS server responsible for managing entries within the Kubernetes installation, updates the video post-processing service IP endpoint, thus making the service available again.

\section{Evaluation and results}\label{ch:eval}

\subsection{End-to-End Latency}

\begin{figure}
  \centering
  \includegraphics[width=0.45\textwidth]{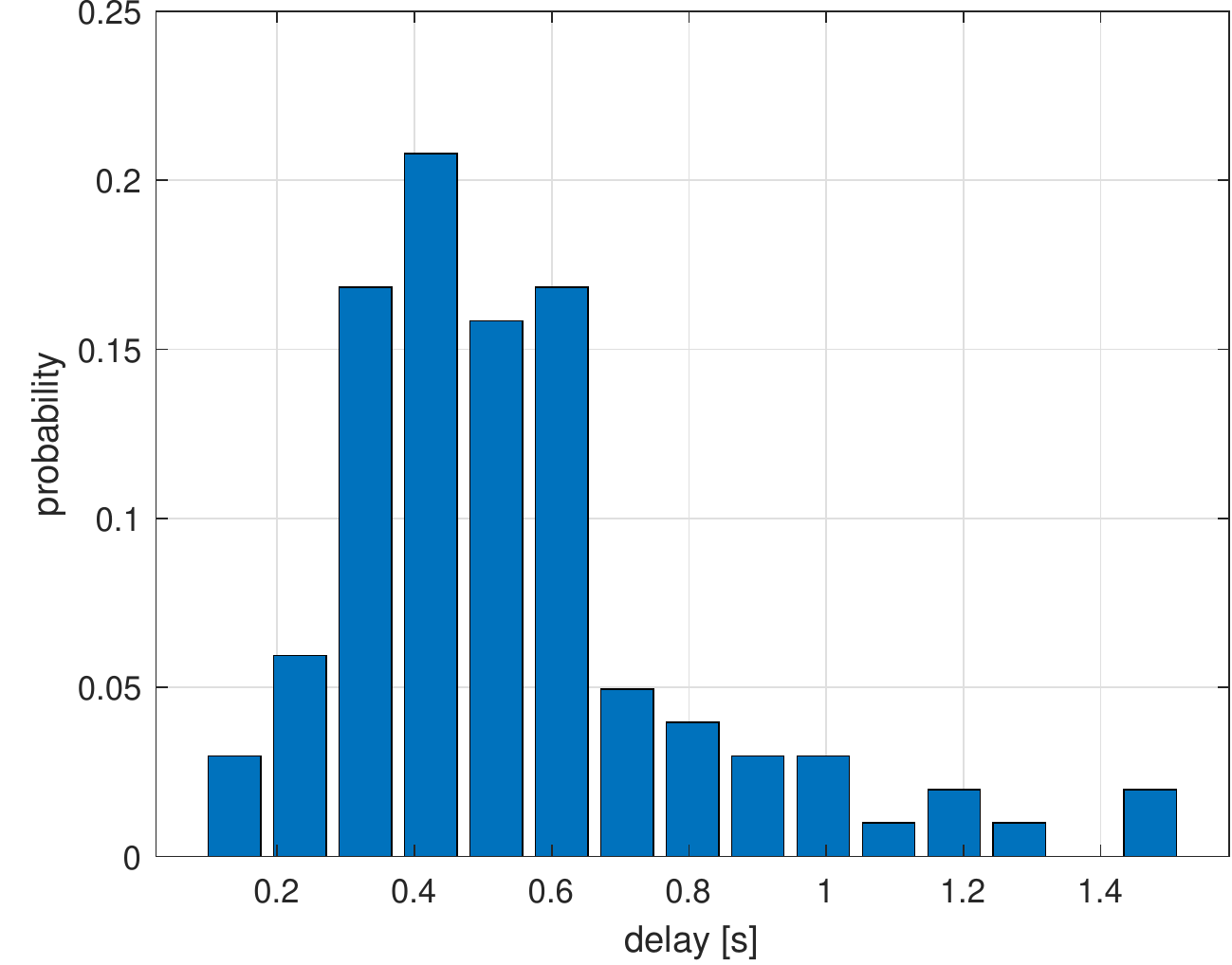}
  \caption{End-to-end latency readings (n=100) visualized as histogram based on empirical cumulative distribution function.}
  \label{fig:e2e}
\end{figure}

The \gls{E2E} latency was measured by holding a stopwatch in front of the \gls{UE} camera on the client application. Now, we took screenshots of the \gls{UE} application, which showed both the live camera feed side-by-side with the post-processed one. Thus, we could count the delta of the stopwatch times seen on the screenshot, resulting in \gls{E2E} latency reading. We measured the end-to-end latency 95\% percentile to be $1.06$s, with minimum and a maximum delay of $0.09$ s and $1.52$s, and mean being $0.55$s, as further demonstrated in Fig. \ref{fig:e2e}.

\subsection{Energy Draw}

Energy draw was measured using a power meter. The power draw was measured from Nvidia Jetson TX2 \gls{SoC} device. The Jetson's hardware architecture is based on ARM; thus, the hardware is similar to those of mobile phones. We measured the Jetson drawing $2.1$ watts on idle and $6.5$ watts while encoding video. Here, we propose that by offloading the video post-processing to the edge, we can reduce power consumption to $1/3$ of what would be needed if the same workload would be done on-device.

\subsection{Application Migration Latency}\label{ss:aml}

Application migration latency was measured from Kubernetes logs. Here, the delta is counted between the time the pod was killed to when the pod self-reported it to be started from its logs. Hence, the factors which count into the migration latency are many and convoluted: e.g., (1) the hardware performance of the Kubernetes scheduler, i.e., the primary host and the secondary nodes, (2) the network fabric, here \verb|flannel| used between Kubernetes primary host and the secondary nodes, (3) the container virtualization method used, e.g., \verb|docker|, which we used, vs. \verb|runc|, vs. \verb|containerd|, vs. et al., (4) the application initialization time, much dependent on programming language, here, Go. While much-related work exists for dissecting the factors and approaches to minimize the latency in each category, we provide our benchmarks as ballpark value. It is part of our on-going work to study approaches to reduce these latency factors towards current and future cellular \gls{URLLC} networks. We measured the application migration latency 95\% percentile to be $6.805$s, with minimum and a maximum delay of $2.730$ s and $7.480$s, and mean being $4.450$s. 

\subsection{Radio Signal Strengths}

\begin{figure*}
    \includegraphics[width=0.99\textwidth]{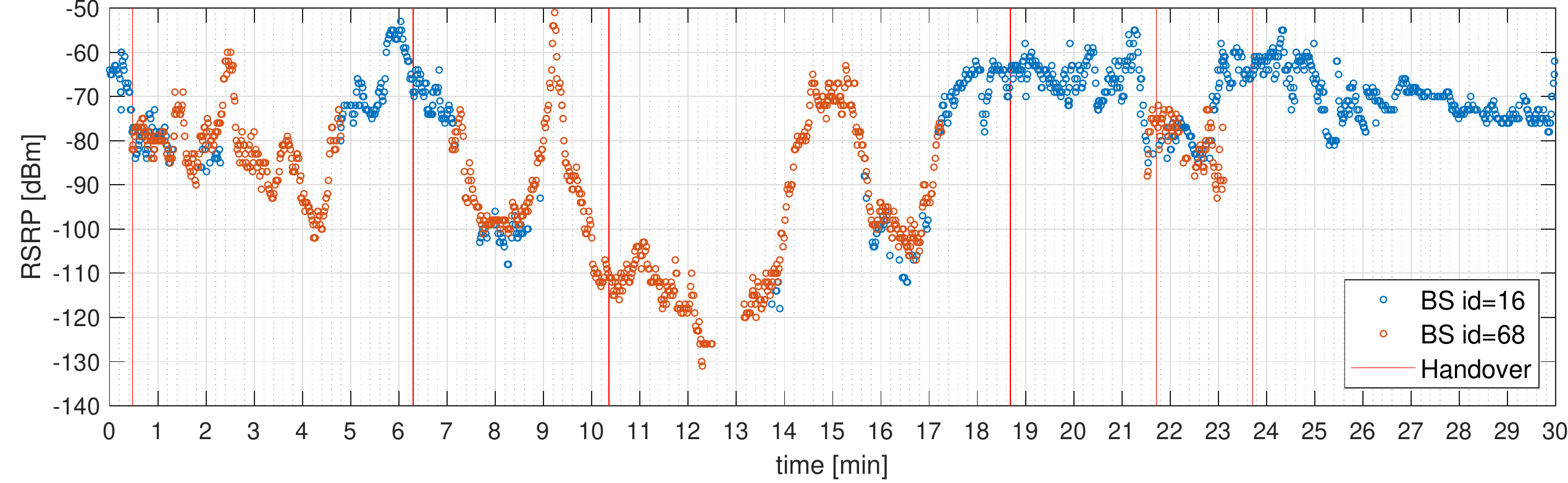}
  \centering
  \caption{RSRP readings. Here, $-70$ or lower indicates an excellent signal, whereas $-110$ or less indicates little to no signal.}
  \label{fig:rsrp}
\end{figure*}

\begin{figure*}
    \includegraphics[width=0.99\textwidth]{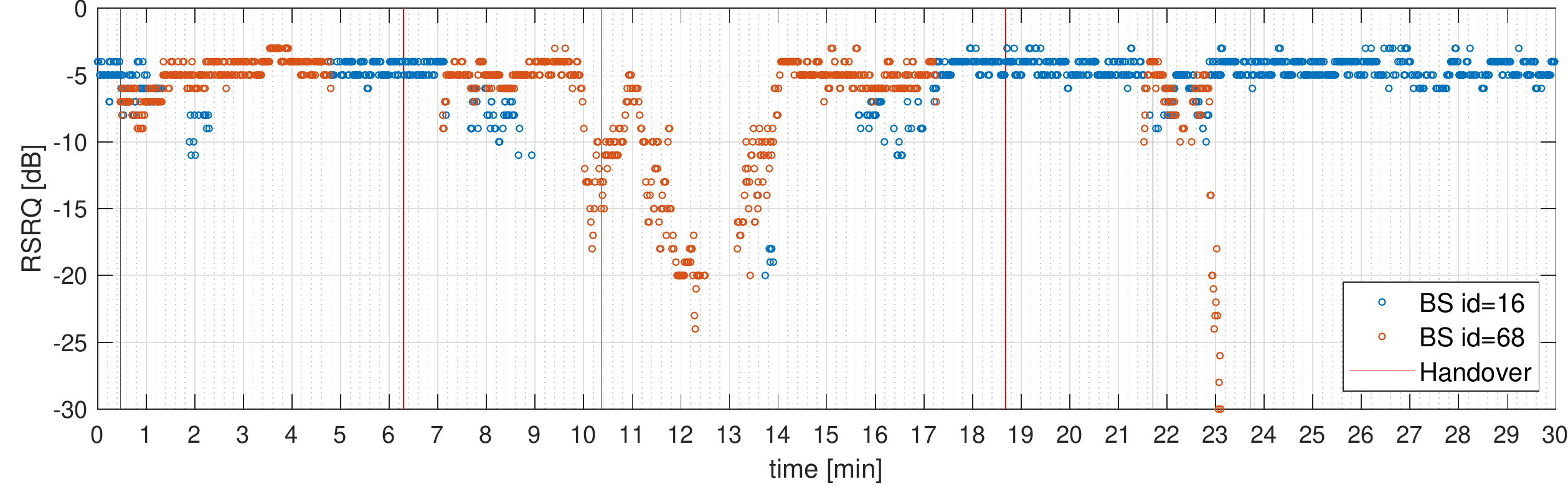}
  \centering
  \caption{RSRQ readings. Here, quality of around -3 indicates an excellent signal, whereas $-16$ or lower as unusable one. }
  \label{fig:rsrq}
\end{figure*}

 For defining the handovers, we recorded the timestamp when the \gls{BS} received an \verb|End Marker| from the \gls{EPC}. \gls{RSS} readings we measured using proprietary software called \textit{Nemo Handy Handheld Measurement Solution} from Keysight Technologies Inc. The software was installed on a Samsung Galaxy S7 phone. Here, the software was used to record \gls{RSRP} (Fig. \ref{fig:rsrp}) and \gls{RSRQ} (Fig. \ref{fig:rsrq}). The idea here was to provide datasets and reference values from which future work towards a pre-emptive \gls{MEC} application migration model could base on. To elaborate, the \gls{MEC} scheduler would use machine learning to migrate and initialize applications between \gls{MEC} hosts ahead of time. This would result in a zero-downtime \gls{MEC} application switch. Now, the \gls{MEC}-dependent end-user application, such as the one proposed in Section \ref{ch:design}, would no longer be prone to downtime as measured in Section\ref{ss:aml}. We envision such model being one practical approach towards addressing problems in cellular \gls{URLLC} application such as autonomous cars, in which the system is at least partly dependent on information coming from the edge. We note this model likely requires both the little varying RSRQ readings and the more varying RSRP readings.

Further, for a scalable system, these values should be collected from the \glspl{BS} instead of individual UEs. However, proprietary \gls{BS} software might make it close to impossible to apply these changes without the cooperation of the hardware providers. This problem could be addressed either by open-source \gls{BS} software, or standardization, such as \glspl{API}, which facilitate software approaches leveraging RAN data analytics for intelligent radio resource optimization.

\section{Conclusion}
\label{ch:concl}

In this research, we studied integrating MEC application migration to S1 radio handovers. In particular, practical concerns of integrating open-source container orchestrator system was addressed, and the constraints under which \gls{UE} might depend on MEC resources. We presented a MEC application which used the MEC resources to post-processes video coming from \gls{UE}. The application was deployed as a Kubernetes service to an existing open-source based RAN at the University of Oulu. The application was successfully deployed and evaluated in a real-world environment. We see worrisome results regarding using existing container orchestration systems, here Kubernetes, for reactively moving MEC applications closer to UEs. Here, we experienced mean container migration time of $4.450$s, suggesting that new approaches to application migration may be needed in future MEC-enabled cellular networks. More generally, we envision that unless these challenges are addressed, applications relying on MEC or fog computing may have a bad quality of service, as the service might experience severe downtime when applications need to be moved between MEC servers for, e.g., better latency. To address these challenges, we gather RSS readings for future research, in which machine learning could be used to pre-emptively move containers ahead-of-time and, as such, address the migration times. Such work is part of our continued studies.

\section{Acknowledgement}

This research is financially supported by Academy of Finland 6Genesis Flagship (grant 318927) and by the AI Enhanced Mobile Edge Computing project, funded by the Future Makers program of Jane and Aatos Erkko Foundation and Technology Industries of Finland Centennial Foundation.

\bibliographystyle{IEEEtran}
\bibliography{IEEEabrv,main}

\end{document}